\documentclass[manuscript]{aastex}

\usepackage{natbib}
\shorttitle{Large-Scale Magnetic Fields and Mass of Galaxies}
\shortauthors{Tabatabaei et al.}
\begin{document}

\title{AN EMPIRICAL RELATION BETWEEN THE LARGE-SCALE MAGNETIC FIELD AND THE DYNAMICAL MASS IN GALAXIES}

\author{F.\,S. Tabatabaei, T.\,P.\,K. Martinsson, J.~H.~Knapen, J.~E.~ Beckman\altaffilmark{1}}
\affil{Instituto de Astrof\'{i}sica de Canarias, V\'{i}a L\'{a}ctea S/N, E-38205 La Laguna, Spain}
\affil{Departamento de Astrof\'{i}sica, Universidad de La Laguna, E-38206 La Laguna, Spain}
\email{ftaba@iac.es}

\author{B.~Koribalski}
\affil{CSIRO Astronomy and Space Science, Australia Telescope National Facility, Epping, NSW 1710, Australia}

\and

\author{B.~G.~Elmegreen}
\affil{IBM T.~J.~Watson Research Center, 1101 Kitchawan Road, Yorktown Heights, NY 10598, USA}

\altaffiltext{1}{Consejo Superior de Investigaciones Cient\`ificas}

\begin{abstract}

The origin and evolution of cosmic magnetic fields as well as the influence of the magnetic fields on the evolution of galaxies are unknown. Though not without challenges, the dynamo theory can explain the large-scale coherent magnetic fields which govern galaxies, but observational evidence for the theory is so far very scarce. Putting together the available data of {non-interacting, non-cluster} galaxies with known large-scale magnetic fields, we find a tight correlation between the integrated polarized flux density, $S_{\rm PI}$,  and the rotation speed, $v_{\rm rot}$, of galaxies. This leads to {an almost} linear correlation between the large-scale magnetic field  ${\bar{B}}$ and $v_{\rm rot}$, assuming that the number of cosmic ray electrons is proportional to the star formation rate, {and a super-linear correlation assuming equipartition between magnetic fields and cosmic rays}. This correlation cannot be {attributed to an active linear $\alpha$-$\Omega$ dynamo}, as no correlation holds with {global} shear or angular speed. {It} indicates instead a coupling between the large-scale magnetic field and the dynamical mass of the galaxies, ${\bar{B}}\sim\,M_{\rm dyn}^{0.25-0.4}$. {Hence, faster rotating and/or more massive galaxies have stronger large-scale magnetic fields. {The observed ${\bar{B}}-v_{\rm rot}$ correlation shows that the anisotropic turbulent magnetic field dominates ${\bar{B}}$ in fast rotating galaxies as the turbulent magnetic field, coupled with gas, is enhanced and ordered due to the strong gas compression and/or {local} shear in these systems}.  {This study supports an stationary condition {for} the large-scale magnetic field as long as the dynamical mass of galaxies is constant.} }

\end{abstract}


\section{INTRODUCTION}
Magnetic fields are present on all scales in the universe from planets and stars to galaxies and galaxy clusters, and even at high redshifts.
They are important for the continuation of life on the Earth, the onset of star formation, the order of the interstellar medium, and the evolution of galaxies \citep[][]{bec13}. Hence, understanding the Universe without understanding magnetic fields is impossible. 

The most widely accepted theory to explain the magnetic fields on stars and planets is the dynamo theory. This describes the process through which a rotating, convecting, and electrically conducting fluid can maintain a magnetic field over astronomical timescales \citep{ste69}. A similar process can also explain the large-scale coherent magnetic fields in galaxies \citep[see][and references therein]{wid02}.   It is assumed that such fields arise from the combined action of {helical turbulence and differential rotation}, a process known as the {$\alpha$-$\Omega$ dynamo}. 
%{  Though there are different  timescale for field amplification in the standard $\alpha\omega$-dynamo may be too long, however, to explain the { large-scale coherent} fields observed in very young galaxies. This could be explained in other dynamo models such as the Parker instability in which the alpha effect is controlled by relativistic cosmic rays produced in young massive stars and supernovae which force loops of magnetic field into the galactic halo. The differential rotation and Coriolis effect further act upon the poloidal loops and amplify the large scale magnetic fields on short timescales of about 200\,Myr \citep{Parker}. 
While a number of fundamental questions concerning the nature of the galactic dynamo remain unanswered, so far no observational evidence for the effect of galaxy rotation on the large-scale magnetic field has been found. This motivated our currently reported investigation of a possible connection between the tracers of the large-scale magnetic field strength and the rotation of galaxies. Finding such a correlation observationally is not necessarily straightforward due to the possible dilution by galaxy-galaxy interactions and environmental effects which influence the rotation curves and possibly the magnetic fields. Such disturbing effects had to be taken into account when selecting the sample. 

The present study is based on a careful measurement of the galaxy rotation speed as well as the model-free tracer of the large-scale magnetic field strength, the integrated polarized flux density ($S_{\rm PI}$).  We introduce the sample and the data in Sect.~2, and describe the $v_{\rm rot}$ measurements in Sect.~ 3. We investigate the possible correlations in Sect.~4 and discuss and summarize the results in Sect. 5 and 6, respectively. 
\section{GALAXY SAMPLE}

Not many galaxies are found in the literature with known polarized intensity measurements. Table~1 shows the 4.8\,GHz integrated polarized intensity measurements for a sample of nearby galaxies, including Local Group galaxies and barred galaxies from \citet{bec02}. To minimize environmental effects, we excluded galaxies in the Virgo and Ursa Major clusters, and those known to be in interacting systems. Measurements with poor signal-to-noise ratio ($\leq$ 1) were also omitted. 
We included the Local Group dwarf galaxies from \citet{chy11}, as they also show large-scale coherent magnetic fields. Galaxies which fit our selection limits, i.e., with known large-scale magnetic fields and minimum environmental disturbances are listed in Table~1. {Each galaxy was observed in polarized light at a linear resolution $d$ smaller than the galaxy optical size (in the sample, $d<0.25\times$\,$R_{25}$, with $R_{25}$ the optical radius). Thus, the large-scale magnetic field with a coherent length $l$ \citep[$\simeq$ {half a galaxy size}, e.g.,][]{flt10} is resolved in all galaxies and the beamwidth depolarization is small ($\sim d^2/l^2 \leq6\%$ in the sample).}    
%
%{  In this sample, galaxies each galaxy was observed with a beamwidth ($\equiv bw\simeq$ angular resolution $\times$ galaxy distance)  smaller than the optical radius the coherent length of the large-scale magnetic field $l$ \citep[$\simeq$ galaxy size, e.g.,][]{flt10}, as  $bw<0.25\times$\,optical radius ($R_{25}$). This ensures us that the observed polarized light is not significantly reduced  due to the beamwidth depolarization ($\sim bw^2/R_{25}^2 \leq6\%$ in the sample). }    

\section{GALAXY ROTATION SPEEDS} 

The adopted rotation speed $v_{\rm rot}$ of the sample galaxies was derived in one of three ways, namely by I) averaging the rotation speed over the flat part of the rotation curves, II) using the corrected $W_{20}$ measurements of HIPASS data,
or III) taking values directly from the literature.

In case I, 10 galaxies had rotation curves derived by \citet{sof99}\footnote{http://www.ioa.s.u-tokyo.ac.jp/$\sim$sofue/RC99/rc99.htm}; these are indicated with `S' in Table~1. The measurements in the `flat part' of the rotation curve were averaged to calculate $v_{\rm rot}$ (see the radial range for individual galaxies in Table~1). The errors were estimated assuming 10 km/s measurement errors \citep{sof97} and 3 degree errors on the inclination. Five galaxies (with reference indicators `A', `B', `BA', `C', `CS') were taken from other papers (see references in Table~1). For the SMC, $v_{\rm rot}$ was derived by a weighted average in the flat part of the rotation curve presented in \citet{bek09}, assuming uncertainties of 6\,km/s in velocity and 5\,degrees in inclination. For NGC~3359, in a similar way, we read the 13 outer data points of the  rotation curve in Figure~11 of \citet{bal83}, excluding the outermost point. For the remaining three galaxies, $v_{\rm rot}$ was calculated in the same way, but with measurements and errors given in their reference papers.

In case II, we derived $v_{\rm rot}$ from HI $W_{20}$ measurements for eight
galaxies (indicated by `W' in Table~1).
The $W_{20}$ measurements were taken from \citet{kor04}  
and have been corrected for instrumental effects, internal turbulent
motion and inclination following \citet{mar15}, assuming a gas velocity
dispersion $\sigma_{\rm HI}=10$\,km/s and using the HIPASS spectral resolution of
18\,km/s.
The measurement errors on $W_{20}$ were estimated to be three times the
error on the systemic velocity as suggested by \citet{kor04}, and the error
on inclination was assumed to be three degrees for all galaxies.

For three galaxies, $v_{\rm rot}$ is taken from the literature \citep[][with superscript `M' in Table~1]{mat98}, only correcting for the inclination (case III).

\section{A TIGHT CORRELATION}

The polarized intensity (PI) provides a measure of the large-scale magnetic field in galaxies. It is defined through the integral over the path length $L$ in the emitting medium

\begin{equation}
{\rm PI} = K \int_L n_{\rm cr} \bar{B}^2 {\rm d}l,   
\end{equation}

with $n_{\rm cr}$ the cosmic ray electron number density, $\bar{B}$ the large-scale transverse magnetic field strength averaged over the line of sight, and $K$ a dimensional constant \citep{bec03}. %(+ anisotropic random field)
The above definition is similar to that of the total intensity of the radio continuum emission, 

\begin{equation}
{\rm I} = K \int_L n_{\rm cr} B^2 {\rm d}l,   
\end{equation}

with $B$ the total transverse magnetic field strength which includes both the large-scale uniform field and the local turbulent field,  $\langle B^2 \rangle=\bar{B}^2+\langle b^2 \rangle$, where   $\langle b \rangle$ is the turbulent magnetic field strength. 
Observationally, the total intensity ${\rm I}$ is just a measure of the Stokes $I$ parameter, while the linearly polarized intensity PI is a measure of the Stokes $Q$ and $U$ parameters,  

\begin{equation}
{\rm PI} = \sqrt{Q^2 + U^2}.   
\end{equation}

The integrated flux densities of the total intensity $S_{\rm I}$ and linearly polarized intensity $S_{\rm PI}$ (in units of Jansky) are obtained {by integrating the ${\rm I}$ and PI  maps (in Jansky/beam) over the area $A$, 
\begin{equation}
S_{\rm PI}=\int {\rm PI}\,\,\, {\rm d}A = K \int \int_L n_{\rm cr}\,\, \bar{B}^2\,\,\, \,{\rm d}l\,\, {\rm d}A,
\end{equation}

\begin{equation}
S_{\rm I} =  \int {\rm I}\,\, {\rm d}A = K \int \int_L n_{\rm cr}\,\, {B}^2\,\,\,\, {\rm d}l\,\, {\rm d}A.
\end{equation}
}

\begin{figure}
\begin{center}
\resizebox{12cm}{!}{\includegraphics*{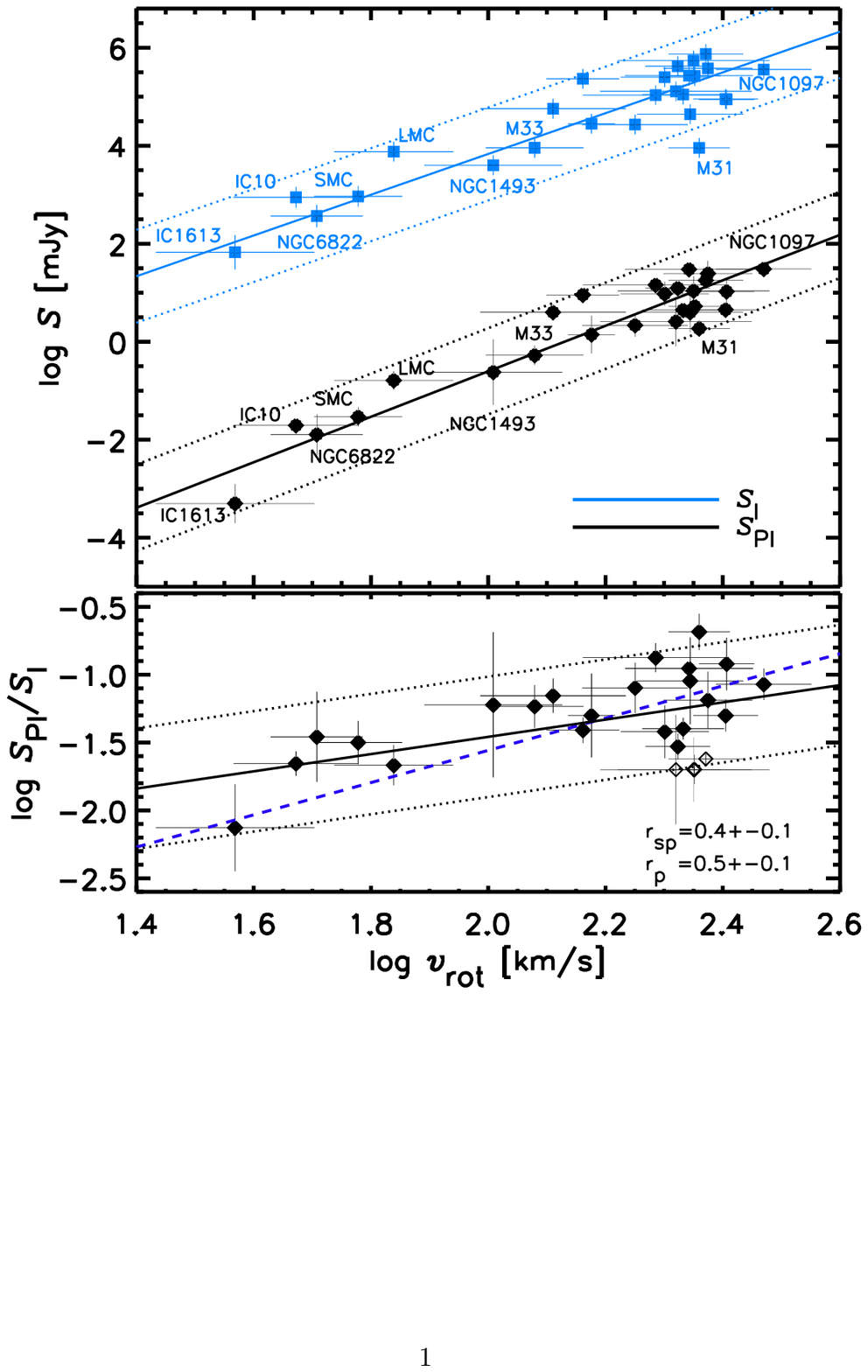}}
\caption[]{{\it Top}: integrated total (top, blue squares) and polarized flux densities (bottom, black dots) projected at a distance of 10\,Mpc vs. the rotation speed of galaxies, $v_{\rm rot}$. {\it Bottom}: fractional polarization $S_{\rm PI}/S_{\rm I}$ against $v_{\rm rot}$. The lines show the least square fits (solid) and their 2$\sigma$ scatter (dotted). In the bottom panel, the dashed line  shows the bisector fit and open symbols indicate NGC~7552, NGC~1365, NGC~5643 (Sect.~5.1).}
\label{fig:fig1}
\end{center}
\end{figure}

For the selected sample of 26 galaxies, we collected $S_{\rm I}$ and $S_{\rm PI}$ at 4.8\,GHz from the literature. Table~1 shows those fluxes, projected to a common distance of 10\,Mpc. Plotting $S_{\rm PI}$ vs. $v_{\rm rot}$ we find a tight correlation (Fig.~\ref{fig:fig1}). The Pearson correlation coefficient is r$_{\rm p}= 0.94\pm0.04$, {and the Spearman rank coefficient r$_{\rm sp}=\,0.8\pm \,0.0$}. The correlation is even tighter than that between $S_{\rm I}$ and $v_{\rm rot}$ {with a rank r$_{\rm sp}=\,0.7\pm \,0.0$} (same figure). A linear fit in logarithmic plane leads to a slope of $4.6\pm0.3$, slightly steeper than that found between $S_{\rm I}$ and $v_{\rm rot}$ ($4.0\pm0.3$). {The ratio of the integrated polarized to total flux density, $S_{\rm PI}$/$S_{\rm I}$, is not constant and increases with $v_{\rm rot}$ {with a least square fit slope of $0.6\pm 0.2$ and a bisector fit slope of $1.2\pm 0.3$}.}

\section{DISCUSSION}
\begin{figure*}
\begin{center}
\resizebox{\hsize}{!}{\includegraphics*{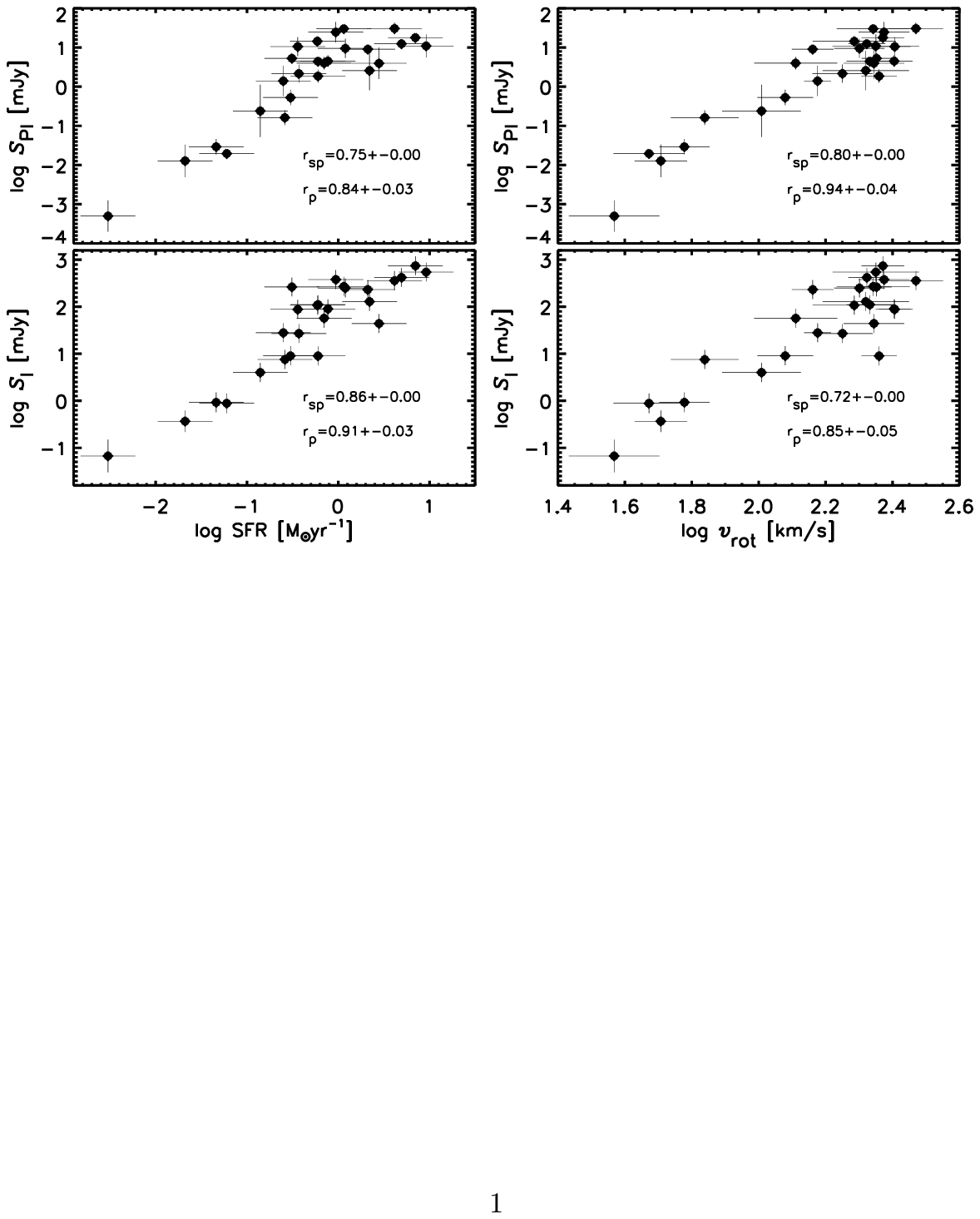}}
\caption[]{Integrated polarized ($S_{\rm PI}$, top) and total  flux densities ($S_{\rm I}$, bottom) against SFR (left panels). {The horizontal error bars show a 25\% uncertainty.} Shown for comparison are $S_{\rm PI}$ and $S_{\rm I}$ against $v_{\rm rot}$ (right panels).}
\label{fig:fig2}
\end{center}
\end{figure*}

\begin{figure}
\begin{center}
\resizebox{8cm}{!}{\includegraphics*{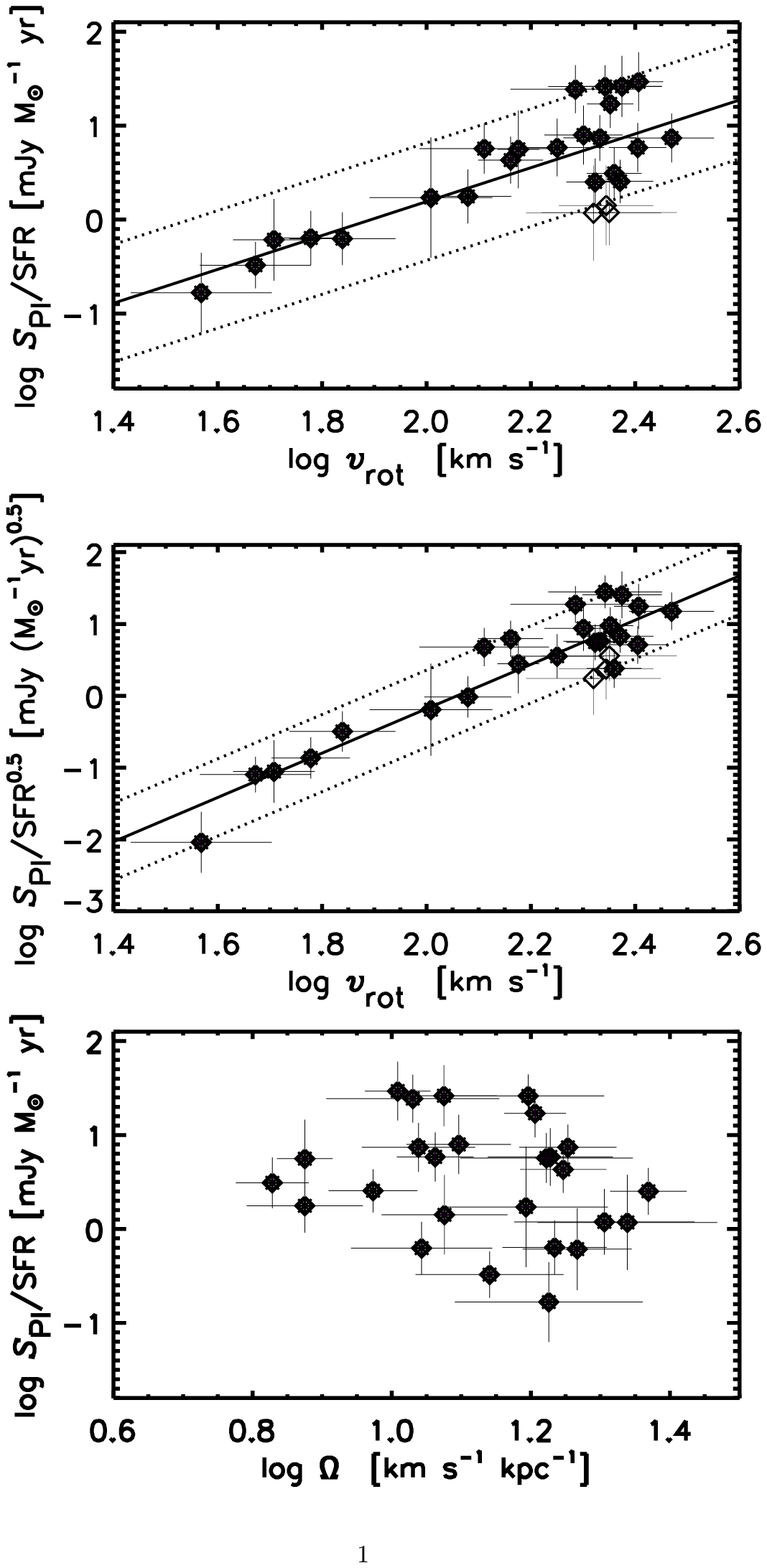}}
\caption[]{{{\it Top}: ratio of the integrated polarized flux density to the SFR vs. the rotation speed.  {\it Middle}: ratio of the integrated polarized flux density to the SFR$^{0.5}$ vs. the rotation speed. Open symbols indicate NGC~7552, NGC~1365, and NGC~5643 (see text).  {\it Bottom}: ratio of the integrated polarized flux density to the SFR vs. the angular speed. }} %The lines show the ordinary least square fit (solid) with a slope of $1.8\pm0.3$  and its 2\,$\sigma$ scatter (dashed).
\label{fig:fig3}
\end{center}
\end{figure}

{Taking into account the well-known correlation between {the total} star formation rate (SFR) and radio luminosity of galaxies \citep[][]{cond92} and the correlation between SFR and stellar mass \citep[e.g.][]{spar15} which scales with $v_{\rm rot}$ according to the Tully-Fisher relation \citep{Tul77}, are the observed correlations $S_{\rm I}-v_{\rm rot}$ and $S_{\rm PI}-v_{\rm rot}$ (Fig.~\ref{fig:fig1}) merely secondary correlations which represent a more direct relation between SFR and $v_{\rm rot}$?}   We estimated  SFR by combining the GALEX far-UV (FUV) \citep{lee11,cor12}  and the IRAS 25$\mu$m fluxes \citep{abr15} and following the calibration relation given by \citet{hao11} and \citet{ken09}:
\begin{eqnarray}
\frac{\rm SFR}{\rm M_{\odot}\,yr^{-1}}\, & = & \,10^{-43.35}\,\left. [\frac{\nu L_{\nu}({\rm FUV})}{\rm erg\,s^{-1}} + \,\, \right. \nonumber \\
& &\left. 3.89\,\times \frac{\nu L_{\nu}({\rm 25\mu m})}{\rm erg\,s^{-1}}] \right.
\nonumber
\end{eqnarray}
In a few cases with no GALEX data, only the IR data were used to estimate the SFR following \citet{ken12}:   
$$ \frac{\rm SFR}{\rm M_{\odot}\,yr^{-1}}= 10^{-42.69}\,\,\frac{\nu L_{\nu}({\rm 24\mu m})}{\rm erg\,s^{-1}}.$$
{Table~1 lists the SFR values obtained. We address the question raised by comparing the correlations of $S_{\rm I}$ and $S_{\rm PI}$ with SFR and $v_{\rm rot}$.  We found that indeed SFR is correlated with $v_{\rm rot}$ with r$_{\rm sp}=0.67$,  similar to the $S_{\rm I}-v_{\rm rot}$ correlation (r$_{\rm sp}=0.72$), indicating that the SFR could be the main cause of the observed $S_{\rm I}-v_{\rm rot}$ correlation. This is shown better by a tighter correlation of $S_{\rm I}$ with SFR (r$_{\rm sp}\sim0.9$) than $v_{\rm rot}$ (r$_{\rm sp}\sim0.7$, Fig.~\ref{fig:fig2}).  

The ${\rm SFR}-v_{\rm rot}$ correlation is, however, {not as tight as} the $S_{\rm PI}-v_{\rm rot}$ correlation (r$_{\rm sp}=0.80$). Hence, the correlation between SFR and $S_{\rm PI}$ (Fig.~\ref{fig:fig2}) {may not} fully explain the tighter $S_{\rm PI}-v_{\rm rot}$ correlation. A correlation is expected between the integrated polarized flux density $S_{\rm PI}$ and SFR due to cosmic rays as their number $N_{\rm cr}$ increases  with star formation activity \citep[e.g.][]{mur08}. Otherwise, no correlation is expected between the large-scale field ${\rm \bar{B}}$ {(traced by PI)} and SFR, while the turbulent (and total) magnetic field positively correlate with SFR \citep{tab013a,tab013b}. This could also explain that the $S_{\rm I}-{\rm SFR}$ correlation is tighter than the $S_{\rm PI}-{\rm SFR}$ correlation (Fig.~\ref{fig:fig2}). 
{We note that SFR is not a directly measured quantity. Potential systematic effects in calibration, like a dependence of the calibration factors on the galaxy properties, could increase  uncertainties in SFR.}

\subsection{The Large-Scale Magnetic Field}
Equation\,(4) can be written as $S_{\rm PI}\sim \langle n_{\rm cr} \rangle\, \bar{B}^2\, {V}$, with $V$ the integration volume. The ratio $S_{\rm PI}/\langle n_{\rm cr}\rangle$ is a measure of the energy of the large-scale magnetic field, E($\bar{B}$)\,=\,${\bar{B}}^2 V/{8\pi}$. Studying the corresponding energy density and the large-scale magnetic field strength is then possible via $S_{\rm PI}/N_{\rm{cr}}\sim \bar{B}^2$ with $N_{\rm cr}= \langle n_{\rm cr}\rangle V$. The SFR can be taken as a proxy for $N_{\rm cr}$ (Sect.~5), and hence  $S_{\rm PI}/{\rm SFR}\,\sim\, \bar{B}^2$. 
Fig.~\ref{fig:fig3}-top shows that  $S_{\rm PI}/{\rm SFR}$ is  correlated with $v_{\rm rot}^{1.8\pm0.3}$ with  r$_{\rm p}=\,0.8\pm0.1$ and r$_{\rm sp}=\,0.7\pm \,0.0$. Thus, $\bar{B}^2\sim\,S_{\rm PI}/{\rm SFR}\,\sim\, v_{\rm rot}^2$ or $\bar{B} \sim v_{\rm rot}$, the large-scale ordered field is linearly proportional to the rotation speed of the galaxies. 

{The relation between $N_{\rm cr}$ and SFR differs if an equipartition between the energy densities of the cosmic ray electrons and the total magnetic field holds. The global equipartition condition leads to $N_{\rm cr}\,\sim\,B^2$, with $B$ the mean total magnetic field strength in each galaxy. Independent observations show a global correlation between $B$ and SFR in galaxies, $B\sim {\rm SFR}^{0.25-0.3}$  \citep{chy11,hee14}. Thus, $N_{\rm cr}$ changes with SFR as $N_{\rm cr}\,\sim {\rm SFR}^{0.5-0.6}$. It then follows that $\bar{B}^2\sim\,S_{\rm PI}/N_{\rm cr}\,\sim S_{\rm PI}/{\rm SFR}^{(0.5-0.6)}$, which is correlated to  $\sim v_{\rm rot}^{3.0\pm0.3}$ with r$_{\rm p}=\,0.9\pm0.1$ (Fig.~\ref{fig:fig3}-middle). Hence, a super-linear correlation holds between $\bar{B}$ and $v_{\rm rot}$:  $\bar{B}\sim v_{\rm rot}^{1.5}$.}

We note that the scatter at the high-$v_{\rm rot}$ end in Fig.~\ref{fig:fig3}- top is due to the galaxies with the highest SFR, NGC~7552 and NGC~1365, as well as the Seyfert galaxy NGC~5643. {The large turbulence due to high SF and AGN activities in these galaxies could enhance the generation of turbulent fields by the small-scale dynamo. It could also} cause depolarization due to internal Faraday dispersion \citep{sok98} reducing the observed PI and $S_{\rm PI}$ and shifting the $S_{\rm PI}/{\rm SFR}$ ratio to relatively low values.

%
%\citep{Klein}

\subsection{Tracing the Dynamo Effect}

Could the observed $\bar{B}-v_{\rm rot}$ {correlations be due to an active} dynamo process? Following the linear mean-field {$\alpha$-$\Omega$ dynamo} theory, the large-scale field is set by the shear, $S=r {\rm d}\Omega/{\rm d}r$, {with $r$ the galactocentric radius and $\Omega$ the angular speed. For a flat rotation curve, the shear is given by } $S=-v_{\rm rot}/r=-\Omega$. Thus, {theoretically,} $\bar{B}$ can be expected to grow proportionally with $\Omega$, and hence the dynamo's appropriate parameter is the angular speed and not the linear speed.  However, we did not find any correlation between $S_{\rm PI}$/SFR and $\Omega=v_{\rm rot}/R_{25}$\footnote{Averaging the actual shear in the flat part of the rotation curves results in $r\sim R_{25}$.} ({see Fig.~\ref{fig:fig3}- bottom}). %{  It is also worth noting that the $\bar{B}-\Omega$ correlation might not even be expected observationally as the strength of the regular field in the linear growth phase could be so weak to produce a detectable polarised radio emission.  }

{A correlation between $\bar{B}$ and $\Omega$ is not expected  theoretically if dynamo quenching and saturation take place \citep{van15}. This saturation occurs when the large-scale magnetic field and turbulent energy densities are similar in the interstellar medium, or because of a balance between Coriolis and Lorentz forces, or due to magnetic helicity \citep[see][and references therein]{wid02,van15}}. In this case, the galactic large-scale dynamos are normally in a non-linear, statistically steady state. As both the quenching and growth of $\bar{B}$ increase with $\Omega$ \citep{cha14}, a correlation between $\bar{B}$ and $\Omega$ is not trivial in the non-linear mean-field dynamo models. {Hence, $\bar{B}$ is not expected to be correlated with {either} $v_{\rm rot}$ {or} $\Omega$ in these models.}
  
%\subsection{Connection to Dynamical Mass}

{The large-scale magnetic field as traced by the polarized emission could be actually dominated by an {\it anisotropic} turbulent field which differs from the field regulated by the dynamo process theoretically. The turbulent magnetic field could become anisotropic and apparently large-scale due to compression and/or local shear by gas streaming velocities \citep[e.g.,][]{lai02}. {Strong anisotropic fields in galaxies were found e.g. in M51 \citep{flt12} and in IC342 \citep{bec15}.} Hence, the $\bar{B}-v_{\rm rot}$ correlation could show that the anisotropic turbulent field dominates in faster rotating galaxies due to higher gas streaming velocities.}  
%This translates to a ${\rm \bar B}-M_{\rm dyn}$ relation which, interestingly, does not differ much from the B-gas coupling in galaxies.  

{On the other hand,} the $\bar{B}-v_{\rm rot}$ correlation could be caused by larger $\bar{B}$ values in galaxies with higher dynamical mass. The dynamical mass inside the optical radius is given by $M_{\rm dyn}\approx R_{25}\,v_{\rm rot}^2/{G}$, with $G$ the gravitational constant. We found that $S_{\rm PI}$/SFR ($\sim {\bar{B}}^2$) increases with $M_{\rm dyn}$ with a slope of $0.5\pm 0.1$, ${\bar{B}}^2\sim M_{\rm dyn}^{0.5\pm0.1}$ (r$_{\rm sp}=\,0.68\pm \,0.00$). {Assuming equipartition, it follows that ${\bar{B}}^2\sim M_{\rm dyn}^{0.8\pm0.1}$.}  
%This implies no further evolution for the strength of the large-scale magnetic field in galaxies as long as their dynamical mass is constant. }

{Earlier attempts to establish a correlation between $\bar{B}$ and the dynamics of galaxies met not much success, although some theoretical works went as far as suggesting an effect of $\bar{B}$ on rotation curves of galaxies \citep{bat07}}. \cite{van15} collected the mean magnetic fields for a sample of 20 nearby galaxies. They however found no correlation between $\bar{B}$  and $v_{\rm rot}$. This, as noted by \cite{van15}, could be due to $a)$ using inconsistent methods/assumptions per galaxy  each referring to a different $\bar{B}$ component and $b)$ inhomogeneous galaxy sample including both interacting and isolated galaxies. We avoided these caveats by using a consistent method for all sources in our sample of the non-interacting/non-cluster galaxies. }

\subsection{Magnetic Energy vs Rotational Energy}
\begin{figure}
\begin{center}
\resizebox{10cm}{!}{\includegraphics*{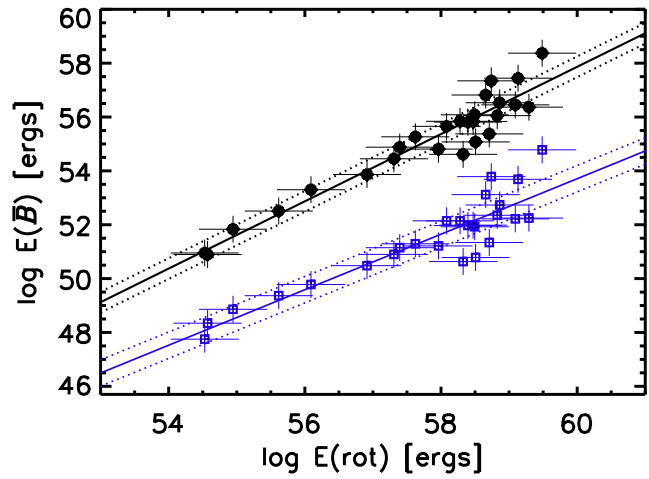}}
\caption[]{Energy of the large-scale magnetic field E(${\bar{B}}$)  vs. the rotational energy E(${\rm rot}$) in ergs  assuming equipartition  (dots). Squares show E(${\bar{B}}$) in arbitrary units vs. E(${\rm rot}$) assuming $N_{\rm cr} \sim {\rm SFR}$. The lines show the ordinary least square fits (solid) and their 5\,$\sigma$ scatter (dotted). The error bars indicate a 50\% uncertainty.  }% 
\label{fig:fig4}
\end{center}
\end{figure}

{The total energy of the large-scale magnetic field scales with $S_{\rm PI}/n_{\rm cr}$ or $S_{\rm PI}\, V/N_{\rm cr}$ (Sect.~1). Assuming $N_{\rm cr}\sim\,{\rm SFR}$, the magnetic energy is
\begin{equation}
{\rm E}_{\bar{B}}\sim\,V\,S_{\rm PI}/{\rm SFR},
\end{equation}
while assuming equipartition leads to 
\begin{equation}
{\rm E}_{\bar{B}}\sim\,V\,S_{\rm PI}/{\rm SFR}^{0.5-0.6}.
\end{equation}
To estimate E$_{\bar{B}}$ using Eqs.\,(6) and (7), the synchrotron radiating volume $V$ is taken to be the volume of the thick disk with the radius of $R_{25}$ and scale hight of 1\,kpc, corrected for inclination. The case of equipartition (Eq.\,(7)) even allows calibration of E$_{\bar{B}}=\,V\,\bar{B}^2/8\pi$ in ergs through 
$$\bar{B}=\,\bar{B}_0 \,\frac{S_{\rm PI}}{S_{\rm PI_0}} \, \left(\frac{\rm SFR_0}{{\rm SFR}}\right)^{0.5-0.6},$$
with $\bar{B}_0$, $S_{\rm PI_0}$ and ${\rm SFR_0}$ the corresponding parameters of a reference galaxy. We took a galaxy residing between the dwarfs and normal spirals, i.e., M33 \citep[$\bar{B}_0=2.5\mu$G,][]{tab08} as the reference galaxy in the sample.

The $v_{\rm rot}$ data allow estimation of the rotational energy of the galaxies, $E_{\rm rot}= \frac{1}{2} I_m \Omega^2\,= \frac{1}{2} I_m \, (v_{\rm rot}/R_{25})^2$, with $I_m\sim M_{\rm dyn}\,R_{25}^2$ the moment of inertia. For disks, $I_m= \frac{1}{2}\,M_{\rm dyn}\, R_{25}^2$ leading to $E_{\rm rot}= \frac{1}{4}\, R_{25}\, v_{\rm rot}^4$. 

Figure.~\ref{fig:fig4} shows the energy of the large-scale magnetic field $E_{\bar{B}}$ against the rotational energy of the galaxies $E_{\rm rot}$. {The slope of the $E_{\bar B}-E_{\rm rot}$ correlation in logarithmic scale is $1.03\pm0.10$  (r$_{\rm p}=0.91\pm0.08$) for $N_{\rm cr}\sim\,{\rm SFR}$, and $1.25\pm0.08$ (r$_{\rm p}=0.96\pm0.06$)
assuming equipartition (the magnetic-to-rotation energy changes as $2\times10^{-4}<E_{\bar B}/E_{\rm rot}<8\times10^{-2}$ with a median value of 0.002).} Therefore, the two different assumptions on $N_{\rm cr}$ do not lead to a significant difference in the $E_{\bar B}-E_{\rm rot}$ relation.     
These correlations show} that galaxies with higher rotational energy have higher energy of the large-scale magnetic field. Considering that the rotational energy is balanced with the gravitational energy and is constant particularly for non-interacting galaxies,  these relations show that  the large-scale magnetic field is almost stationary at the current epoch in each galaxy {(this does not reject a possibly  linear dynamo process in the past).} This supports the theoretical predictions that the present-day large-scale dynamos cannot be in their linear (growing) phase.
%although the galaxies could have experienced a linear dynamo process during their evolution history.  

Considering active star formation as the main source of the turbulence, the ${\rm SFR}-v_{\rm rot}$ correlation implies that the turbulent energy does not change much, as long as the galaxy mass is fixed. Several studies show an equipartition between the turbulent and the {\it total} magnetic field energy densities \citep[e.g.,][]{beck07,bec15,tab08} indicating that the total magnetic field also does not evolve further.

\section{SUMMARY}

To study the relations between galactic magnetic fields and rotation, we selected from the literature observations of the large-scale magnetic field of galaxies which are not in clusters and are not interacting. These show a tight correlation between the polarized flux density and the rotation speed $v_{\rm rot}$ over three orders of magnitude in dynamical mass. {Assuming $N_{\rm cr}\sim\,{\rm SFR}$, a linear correlation is found between the large-scale magnetic field strength ${\bar{B}}$  and $v_{\rm rot}$. 
{This correlation is super-linear (${\bar{B}}\sim\,v_{\rm rot}^{1.5}$) assuming equipartition between the cosmic ray electrons and the total magnetic field.} 
On the other hand, no correlation is found between ${\bar{B}}$ and $\Omega$, expected from the linear {$\alpha$-$\Omega$ dynamo theory}. The ${\bar{B}}-v_{\rm rot}$ correlation suggests that there is a coupling between the large-scale magnetic field and the dynamical mass of galaxies: {${\bar{B}}\sim\,M_{\rm dyn}^{0.25-0.4}$.} Therefore, faster rotating/ more massive  galaxies have stronger magnetic fields than slower rotating/ less massive galaxies. 
The  ${\bar{B}}-v_{\rm rot}$ correlation shows that the anisotropic turbulent field dominates the large-scale field in faster rotating galaxies, perhaps due to high streaming velocities of the gas with which the magnetic field is coupled.   

{A comparison between the magnetic and rotational energies shows that the large-scale magnetic field is stationary and does not evolve further in galaxies with fixed dynamical mass.}  
%A similarly stationary condition exists for the turbulent energy, assuming that it is set by SFR which is correlated with the dynamical mass fixed in the non-interacting galaxies. 

This is the first study showing a statistically meaningful effect of the galaxy rotation/ mass on {the large-scale magnetic fields}.}

\acknowledgments
{We thank the anonymous referee for his/her valuable comments.}
We also thank Anvar Shukurov and Andrew Fletcher for stimulating discussions and {to} John Dickel for providing us with the LMC and SMC data. FST, TPKM, and JHK acknowledge financial support from the Spanish Ministry of Economy and Competitiveness (MINECO) under grant number AYA2013-41243-P. JHK and JEB acknowledge financial support to the DAGAL network from the People Programme (Marie Curie Actions) of the European UnionÕs Seventh Framework Programme FP7/2007-2013/ under REA grant agreement number PITN-GA-2011-289313.

\clearpage
\clearpage
\clearpage
\clearpage

\clearpage
\begin{deluxetable}{ccrrrrrrcrl}
\tabletypesize{\scriptsize}
\rotate
\tablecaption{Properties of the galaxy sample.}
\tablewidth{0pt}
\tablehead{
\colhead{Galaxy} & \colhead{Hubble} & \colhead{Inclination} & \colhead{Distance} & \colhead{$S_{\rm I}$} & \colhead{$S_{\rm PI}$} & \colhead{$v_{\rm rot}$} & \colhead{$R_{\rm min}$} & \colhead{$R_{\rm max}$} & \colhead{SFR} & \colhead{$R_{25}$} \\
\colhead{Name} & \colhead{Type} & \colhead{[$^{\circ}$]} & \colhead{[Mpc]} & \colhead{[mJy]} & \colhead{[mJy]} & \colhead{[km/s]} & \colhead{[kpc]} & \colhead{[kpc]} & \colhead{[M$_{\sun}$/yr]} & \colhead{[kpc]} }
\startdata
IC342  &   SABcd  & 25          &    3.1    &    108\,$\pm$\,9$^{1}$   &    14\,$\pm$\,2$^{1}$ &  193\,$\pm$\,24$^{\rm S}$ & 5& 18& 0.59&11.87\\
NGC6946&  SABcd &   30         &     6.8     &    270\,$\pm$\,55$^{2}$  &    30\,$\pm$\,1.1$^{2}$ & 220\,$\pm$\,24$^{\rm S}$  & 6& 14& 1.15&9.18\\
NGC253 &   SABc       &    78    &3.94        & 420\,$\pm$\,25$^{4}$      &12.4\,$\pm$\,1.5$^{4}$ & 211\,$\pm$\,12$^{\rm S}$& 2& 9& 4.94&13.77\\
NGC3628 & SAb\,pec& 86     & 6.7              &111\,$\pm$\,20$^{5}$       &   4.4\,$\pm$\,0.4$^{5}$   & 215\,$\pm$\,15$^{\rm S}$&2 &12 & 0.6&18.34\\
NGC4565 & SAb   &86       &        13.1       &88\,$\pm$\,10$^{5}$       &  10.6\,$\pm$\,2.5$^{5}$ &  255\,$\pm$\,12$^{\rm S}$ &5 & 25& 0.36&62.61\\
NGC4736&  SABab &    35        &    4.7     &    27\,$\pm$\,6$^{6}$   &    2.16\,$\pm$\,0.50$^{6}$ & 178\,$\pm$\,16$^{\rm S}$ & 3& 7& 0.37&7.91\\  
NGC5907 & Sc  &88             & 16.4        &    90\,$\pm$\,7$^{7}$   &   4.5\,$\pm$\,0.7$^{7}$ & 254\,$\pm$\,14$^{\rm S}$&5 & 22& 0.77&28.14\\
NGC891  & SAb   & 88            &   8.4      & 264\,$\pm$\,45$^{5}$    &5.3\,$\pm$\,0.7$^{5}$   &225\,$\pm$\,10$^{\rm S}$ &4 &14 & 0.31&18.48\\
NGC1097  &    SBbc & 40         & 16          & 359\,$\pm$\,30$^{3}$     & 30.5\,$\pm$\,4.4$^{3}$    &295\,$\pm$\,24$^{\rm S}$& 5& 27 & 4.13&22.35\\  
NGC1365 &    SBb&   46          & 19          & 744\,$\pm$\,47$^{3}$     & 17.9\,$\pm$\,0.9$^{3}$    & 235\,$\pm$\,15$^{\rm S}$ & 15&25 & 7&35.01\\ 
NGC3359 &   SBc& 51             &  11         & 28\,$\pm$\,2$^{3}$       &1.4\,$\pm$\,0.8$^{3}$      & 149\,$\pm$\,6$^{\rm BA}$ &11 & 20& 0.25&20.65\\ 
%NGC3992 &  SBb  &   59          &  15     &  10\,$\pm$\,2     &1.1\,$\pm$\,0.4$^{3}$    & 242\,$\pm$\,5$^{\rm V}$ &... &... &0.12\\ 
NGC1493&   SBc&  27             &     12     &  4\,$\pm$\,1$^{3}$      &0.24\,$\pm$\,0.16$^{3}$    & 102\,$\pm$\,12$^{\rm W}$ &... &... &0.14& 6.54\\
NGC1559 &  SBc&   57            &     15      &232\,$\pm$\,13$^{3}$    & 9.1\,$\pm$\,1.1$^{3}$      & 145\,$\pm$\,9$^{\rm W}$ &... &... &2.1&8.22 \\
NGC1672 &  SBb&  39             &     15      &250\,$\pm$\,14$^{3}$     &9.5\,$\pm$\,2.3$^{3}$       &200\,$\pm$\,15$^{\rm W}$& ...&... & 1.2&16.02\\
NGC3059&  SBc&    27            &     14      & 57\,$\pm$\,4$^{3}$      &3.99\,$\pm$\,0.63$^{3}$    &129\,$\pm$\,16$^{\rm W}$ &... &... &0.7& 7.73\\
NGC5643& SBc &    23           &      14      & 129\,$\pm$\,8$^{3}$      & 2.6\,$\pm$\,1.3$^{3}$    & 209\,$\pm$\,27$^{\rm W}$&... &... &2.2& 9.58\\
NGC7552& SBbc&    28           &      21     & 546\,$\pm$\,87$^{3}$     & 10.9\,$\pm$\,3.2$^{3}$   & 224\,$\pm$\,29$^{\rm W}$ & ...& ...&9.2&11.08\\
NGC1300& SBb&    35            &      20     &  44\,$\pm$\,8$^{3}$     &3.96\,$\pm$\,1.58$^{3}$   & 221\,$\pm$\,20$^{\rm W}$ &... &... &2.8&18.56\\
NGC7479&  SBbc& 45             &      34     &  379\,$\pm$\,46$^{3}$   &24.6\,$\pm$\,6.4$^{3}$    &237\,$\pm$\,18$^{\rm W}$  & ...&... &0.9&19.94\\
%NGC3953 &SBbc&      61          &       15    &  16\,$\pm$\,2     &1.1\,$\pm$\,0.5$^{3}$    &223\,$\pm$\,5$^{\rm V}$ & ...&... &0.3\\
M31    &  SAb  &   77         &  0.69       & 9\,$\pm$\,1$^{8}$       & 1.9\,$\pm$\,0.3$^{8}$       &  229\,$\pm$\,12$^{\rm C}$& 16& 34 &0.6&21.06\\
M33    &   SAcd&     54        &    0.84     &    9.06\,$\pm$\,0.95$^{9}$   &    0.53\,$\pm$\,0.1$^{9}$ & 120\,$\pm$\,10$^{\rm CS}$ &... &... &0.30&8.49\\
LMC    & Irr/SBm &   33        &    0.050    &    7.6\,$\pm$\,1.5$^{10}$&    0.16\,$\pm$\,0.03$^{10}$ & 69\,$\pm$\,7$^{\rm A}$& 2.5& 13& 0.26&4.70\\
SMC    &   Irr  &   40    &     0.060    &    0.9\,$\pm$\,0.2$^{11}$&     0.029\,$\pm$\,0.006$^{11}$ & 59\,$\pm$\,4.5$^{\rm B}$ &... &... &0.046&2.75\\ 
IC10   &   Irr&  40       &  0.66      & 0.88\,$\pm$\, 0.15$^{12}$  & 0.020\,$\pm$\,0.002$^{12}$  & 47\,$\pm$\,5$^{\rm M}$ & 2.68&3.40 & 0.06&0.60\\ 
NGC6822  &    Irr& 67              &  0.50      & 0.37\,$\pm$\, 0.07$^{12}$ & 0.013\,$\pm$\,0.005$^{12}$   &  51\,$\pm$\,4$^{\rm M}$ &... &... & 0.02&1.13\\
IC1613 &    Irr& 35        &  0.73      & 0.07\,$\pm$\,0.01$^{12}$  & 0.0005\,$\pm$\,0.0002$^{12}$ &  37\,$\pm$\,5$^{\rm M}$ &... &... &0.003&1.72\\
\enddata
%% Text for table notes should follow after the \enddata but before
%% the \end{deluxetable}. Make sure there is at least one \tablenotemark
%% in the table for each \tablenotetext.
\tablecomments{The integrated flux densities $S_{\rm I}$ and $S_{\rm PI}$ were measured at 4.8\,GHz and projected to a disctance of 10\,Mpc. The Hubbel type and distance are taken from the NASA Extragalactic Database,  1- \citet{gra88}, 2- \citet{bec96}, 3-\citet{bec02} 4- \citet{hee09}, 5- \citet{dum98}, 6- \citet{chy08}, 7- \citet{dum00},  8- \citet{ber03}, 9- \citet{tab07,tab08}, 10- \citet{dick05}, 11- \citet{dick10}, 12- \citet{chy11},   A- \citet{alv00}, B- \citet{bek09}, BA- \citet{bal83}, C- \citet{car06}, CS- \citet{corb00}, M- \citet{mat98}, S- \citet{sof99}, W- \citet{kor04}. The inclinations were taken from the $v_{\rm rot}$ references.}

\end{deluxetable}

%% If you use the table environment, please indicate horizontal rules using
%% \tableline, not \hline.
%% Do not put multiple tabular environments within a single table.
%% The optional \label should appear inside the \caption command.

\clearpage

%% If the table is more than one page long, the width of the table can vary
%% from page to page when the default \tablewidth is used, as below.  The
%% individual table widths for each page will be written to the log file; a
%% maximum tablewidth for the table can be computed from these values.
%% The \tablewidth argument can then be reset and the file reprocessed, so
%% that the table is of uniform width throughout. Try getting the widths
%% from the log file and changing the \tablewidth parameter to see how
%% adjusting this value affects table formatting.

%% The \dataset{} macro has also been applied to a few of the objects to
%% show how many observations can be tagged in a table.

\end{document}